\documentstyle{article}

\newtheorem{example}{Example}
\newtheorem{theorem}{Theorem}

\begin{document}

\title{The Heisenberg Representation of Quantum Computers}
\author{Daniel Gottesman\thanks{gottesma@t6-serv.lanl.gov}\\
T-6 Group\\Los Alamos National Laboratory\\Los Alamos, NM 87545}
\maketitle

\begin{abstract}  
Since Shor's discovery of an algorithm to factor numbers on
a quantum computer in polynomial time, quantum computation
has become a subject of immense interest.  Unfortunately, one
of the key features of quantum computers --- the difficulty
of describing them on classical computers --- also makes it
difficult to describe and understand precisely what can be
done with them.  A formalism describing the evolution of
operators rather than states has proven extremely fruitful
in understanding an important class of quantum operations.
States used in error correction and certain communication
protocols can be described by their stabilizer, a group
of tensor products of Pauli matrices.  Even this simple
group structure is sufficient to allow a rich range of
quantum effects, although it falls short of the full power
of quantum computation.
\end{abstract}

\section{Introduction}

Computers are physical objects.  While that seems obvious, it has some
profound consequences.  The familiar desktop and laptop computers and
their big cousins, the mainframes, all work in essentially the same
way.  Their components are individual registers called {\em bits},
which can interact through various discrete operations, called {\em
gates}.  The underlying hardware varies somewhat from system to
system, but currently, all bits are made up of fairly large numbers of
atoms.  Consequently, they behave like macroscopic {\em classical}
objects.  Modern computers are digital, so their bits take the
discrete values of 0 and 1 (thus the name ``bit'').

However, suppose instead that we had a computer whose bits were made
up of a small number of atoms, perhaps even just one atom per bit.
Then the bits in the computer would start to behave like quantum
objects instead of classical objects.  Instead of the individual
values 0 and 1, a bit might instead be in some superposition of 0 and
1.  A bit that can be in such a superposition is called a {\em quantum
bit} or {\em qubit}.  If a classical computer has $N$ bits, it has a 
total of $2^N$ possible states.  In contrast, a {\em quantum
computer} with $N$ qubits can be in any superposition of those $2^N$
classical basis states, resulting in an arbitrary state in a
$2^N$-dimensional Hilbert space.  To even describe such a state on a
classical computer would require $2^N - 1$ complex numbers.

Classical algorithms are frequently classified as {\em polynomial} or
{\em exponential} (which actually just means more than polynomial).  A
polynomial algorithm to solve a problem of size $N$ takes $C N^k$
steps (plus lower-order terms), for some constants $C$ and $k$.  Such
an algorithm is considered to be efficient for dealing with large
problems, since the computation cost does not increase {\em too} much
as $N$ increases (although if $C$ and/or $k$ are large, the algorithm
may not be useful for realistic problems).  An exponential time
algorithm, on the other hand, means it will be very difficult to solve
large problems.  Unfortunately, many important problems have no known
polynomial algorithm, forcing us to resort to approximations or
restricting us to small instances of the problem.

As long as we are using fundamentally classical bits, the algorithms
we use will be roughly equivalent.  Perhaps some system uses a factor
of $N$ more or less steps than another, or perhaps $C$ is larger or
smaller, but an exponential algorithm on one classical computer will
translate to an exponential algorithm on any other classical
computer.  However, if we change the underlying laws of physics and
work with quantum bits instead of classical bits, the situation
changes.  Even if there is no polynomial classical algorithm to solve
a problem, there may be a polynomial quantum algorithm.

Perhaps the most dramatic known example of this is Shor's factoring
algorithm~\cite{factor}.  Factoring an $n$-bit number can be very
difficult, particularly if it is the product of two primes of about
the same size.  Despite much work, no polynomial classical algorithm
is known.  In fact, many classical cryptographic protocols, such as
RSA, depend on the fact that factoring is hard to do.

It turns out that factoring a number $N$ is essentially equivalent (up
to some straightforward additional computation) to finding the order
$r$ of a random number $x$ modulo $N$.  On a quantum computer, we can
do this by starting with a superposition $\sum |a\rangle$ and
calculating $x^a \bmod N$ in a second register, producing the state
$\sum |a\rangle |x^a\rangle$.  This state is periodic in $a$, with
period $r$.  Therefore, by performing the discrete Fourier transform
and measuring the result, we can extract $r$ and thus factor $N$.  The
result is a polynomial quantum algorithm where all known classical
algorithms are exponential.

Unfortunately, because quantum computers deal with arbitrary states in
such a large Hilbert space, it can be very difficult to construct and
analyze even relatively small networks of quantum gates.  It turns
out, however, that a restricted class of networks can be comparatively
easily described if we follow the evolution, not of the state of the
quantum computer, but instead of a set of operators that could act on
the computer~\cite{gottesman:FT}.  In some ways, this is analogous to
the standard Heisenberg representation of quantum mechanics, where the
operators evolve in time, as opposed to the Schr\"odinger picture, where
the states evolve.  Similar, but less powerful techniques, have been
used in the NMR community for years under the name ``the product
operator formalism.''  For a strengthening of the product operator
formalism with applications to quantum computing, see~\cite{prodop}.

\section{Basics of the Heisenberg Representation}

Suppose we have a quantum computer in the state $|\psi\rangle$, and we
apply the operator $U$.  Then 
\begin{equation}
U N |\psi\rangle = U N U^\dagger U |\psi\rangle,
\end{equation}
so after the operation, the operator $U N U^\dagger$ acts on states in
just the way the operator $N$ did before the operation.  Therefore,
applying $U$ to the computer transforms an arbitrary operation $N$ by
\begin{equation}
N \rightarrow U N U^\dagger.
\label{trans}
\end{equation}
By following the evolution of a sufficiently large set of $N$'s, we
will be able to completely reconstruct the evolution of the state
vector.  The evolution (\ref{trans}) is linear, so it will be
sufficient to follow a set that spans the set of $n \times n$ matrices
${\cal M}_n$.  

For this set, we choose the {\em Pauli group} {\cal P}, which consists
of $4 \cdot 4^n$ elements.  The Pauli group contains the $4^n$
$n$-qubit tensor products of the identity $I$ and the Pauli matrices
$\sigma_x$, $\sigma_y$, and $\sigma_z$:
\begin{equation}
\sigma_x = \pmatrix{0 & 1 \cr 1 & 0},\ \sigma_y = \pmatrix{0 & -i \cr i
& \ 0},\ \sigma_z = \pmatrix{1 & \ 0 \cr 0 & -1}.
\end{equation}
Note that
\begin{equation}
\sigma_x^2 = \sigma_y^2 = \sigma_z^2 = I
\end{equation}
and
\begin{equation}
\sigma_y = i \sigma_x \sigma_z.
\end{equation}
Therefore, to complete the group, we must allow each of the $4^n$
tensor products to have an overall phase of $\pm 1$ or $\pm i$.
Below, I will write $X,$ $Y,$ and $Z$ instead of $\sigma_x$,
$\sigma_y$, and $\sigma_z$.\footnote{In earlier publications, $Y$ was
instead equal to $-i \sigma_y$.}  The operators $X_i,$ $Y_i,$ and
$Z_i$ are $X,$ $Y,$ and $Z$ acting on the $i$th qubit in the computer.

Furthermore, (\ref{trans}) is a multiplicative group homomorphism:
\begin{equation}
M N \rightarrow U M N U^\dagger = \left( U M U^\dagger \right)
\left( U N U^\dagger \right).
\end{equation}
Therefore, we can deduce the behavior of any element of the Pauli
group by just following the evolution of a generating set.  A
convenient generating set is just $\{X_1, \ldots, X_n, Z_1, \ldots,
Z_n\}$.  To completely specify a general operator, we need only
describe the evolution of $2n$ single-qubit operators.

\section{The Clifford Group}

Now, in general, the transformation (\ref{trans}) could take a Pauli
matrix $N$ to any of a rather large class of unitary operators.  If we
consider arbitrary quantum gates, the description of the
transformation of our generating set will rapidly become unmanageably
large.  Instead, we will consider a restricted class of gates --- the
gates which transform elements of the Pauli group into other elements
of the Pauli group.

The set of operators which leave the group ${\cal P}$ fixed under
conjugation form a group $N({\cal P})$, the {\em normalizer} of ${\cal
P}$ in $U(2^n)$.  $N({\cal P})$ is also called the {\em Clifford
group} ${\cal C}$ for its relationship to the usual Clifford groups
and Clifford algebras.  While ${\cal C}$ is considerably smaller than
the full unitary group $U(2^n)$, it still contains a number of
operators of particular interest.

For instance, the Clifford group contains the single-qubit Hadamard
transform R:
\begin{equation}
R |j\rangle = |0\rangle + (-1)^j |1\rangle,\ R = \frac{1}{\sqrt{2}}
\pmatrix{1 & \ 1 \cr 1 & -1}.
\end{equation}
Another element is the phase gate P:
\begin{equation}
P |j\rangle = i^j |j\rangle,\ P = \pmatrix{1 & 0 \cr 0 & i}.
\end{equation}
It also contains the controlled-NOT (CNOT) gate, also known as the XOR:
\begin{equation}
{\rm CNOT} |j\rangle |k\rangle = |j\rangle |j + k \bmod 2\rangle.
\end{equation}
The first qubit is called the {\em control} qubit and the second is
the {\em target} qubit.

In fact, R, P, and CNOT, applied to arbitrary qubits or pairs of
qubits, generate the full Clifford group ${\cal C}$.  The
transformations these three gates induce on the Pauli matrices are
given in table~\ref{table:Clifford}.
\begin{table}
\centering
\begin{tabular}{lcc}
R & $\begin{array}[t]{c} X \rightarrow Z \\ Z \rightarrow X \end{array}$ & 
\begin{picture}(0,0)(0,10)
\put(0,10){\line(1,0){9}}
\put(9,4){\framebox(12,12){R}}
\put(21,10){\line(1,0){9}}
\end{picture} \\
\\
P & $\begin{array}[t]{c} X \rightarrow Y \\ Z \rightarrow Z \end{array}$
& \begin{picture}(0,0)(0,10)
\put(0,10){\line(1,0){9}}
\put(9,4){\framebox(12,12){P}}
\put(21,10){\line(1,0){9}}
\end{picture} \\
\\
CNOT & $\begin{array}[t]{c} X \otimes I \rightarrow X \otimes X \\
I \otimes X \rightarrow I \otimes X \\
Z \otimes I \rightarrow Z \otimes I \\
I \otimes Z \rightarrow Z \otimes Z \end{array}$ &
\begin{picture}(0,0)(0,30)
\put(0,30){\line(1,0){30}}
\put(0,10){\line(1,0){30}}
\put(15,30){\circle*{4}}
\put(15,30){\line(0,-1){24}}
\put(15,10){\circle{8}}
\end{picture} \\
\\
\multicolumn{2}{l}{\begin{tabular}{l}
Measurement of $A$, plus operation $B$ \\
performed if measurement result is $-1$ \end{tabular}} &
\begin{picture}(0,0)(0,20)
\put(0,30){\line(1,0){9}}
\put(21,30){\line(1,0){9}}
\put(0,10){\line(1,0){9}}
\put(21,10){\line(1,0){9}}

\put(15,30){\circle{12}}
\put(9,24){\makebox(12,12){$A$}}
\put(15,24){\line(0,-1){8}}
\put(9,4){\framebox(12,12){$B$}}
\end{picture} \\
\end{tabular}
\caption{Generators of the Clifford group, and their induced
transformations on the Pauli matrices.}
\label{table:Clifford}
\end{table}
The table also gives the symbols used to represent the Clifford group
operations in gate networks, as well as the symbol for a measurement.

If we apply R or P to qubit number $j$, I will write $R(j)$ or $P(j)$
to represent the gate.  If we apply a CNOT with qubit $j$ as control
and $k$ as target, I will write ${\rm CNOT}(j \rightarrow k)$.

\section{Applying the Heisenberg Representation}

To see how to apply the Heisenberg representation, consider the
following example:

\begin{example}
Alice's quantum computer is working too well.  Instead of performing
single controlled-NOT gates, it does three at a time (figure
\ref{fig:swap}a).  What is it actually doing?
\end{example}
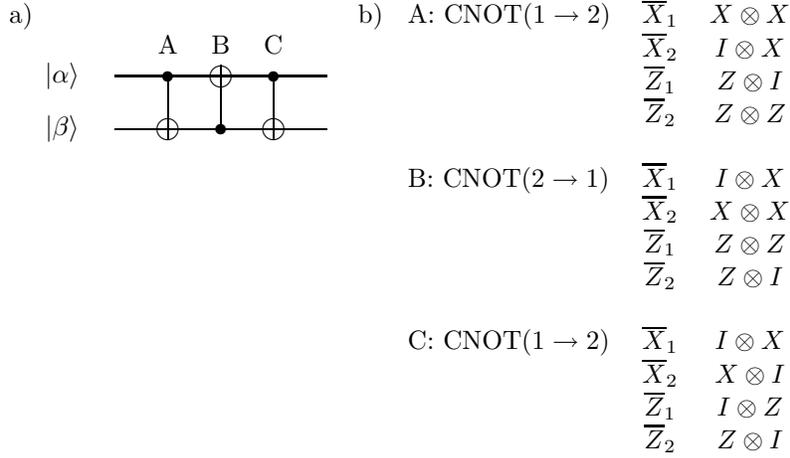
\begin{figure}
\centering
\begin{tabular}{cc}
\begin{picture}(120,10)(0,60)

\put(0,60){a)}
\put(0,34){\makebox(40,12){$|\alpha\rangle$}}
\put(0,14){\makebox(40,12){$|\beta\rangle$}}

\put(40,40){\line(1,0){80}}
\put(40,20){\line(1,0){80}}

\put(60,40){\circle*{4}}
\put(60,40){\line(0,-1){24}}
\put(60,20){\circle{8}}

\put(80,20){\circle*{4}}
\put(80,20){\line(0,1){24}}
\put(80,40){\circle{8}}

\put(100,40){\circle*{4}}
\put(100,40){\line(0,-1){24}}
\put(100,20){\circle{8}}

\put(54,46){\makebox(12,12){A}}
\put(74,46){\makebox(12,12){B}}
\put(94,46){\makebox(12,12){C}}

\end{picture}
&
b) \begin{tabular}[t]{lcc}
A: ${\rm CNOT}(1 \rightarrow 2)$ & $\overline{X}_1$ & $X \otimes X$ \\
& $\overline{X}_2$ & $I \otimes X$ \\
& $\overline{Z}_1$ & $Z \otimes I$ \\
& $\overline{Z}_2$ & $Z \otimes Z$ \\
\\
B: ${\rm CNOT}(2 \rightarrow 1)$ & $\overline{X}_1$ & $I \otimes X$ \\
& $\overline{X}_2$ & $X \otimes X$ \\
& $\overline{Z}_1$ & $Z \otimes Z$ \\
& $\overline{Z}_2$ & $Z \otimes I$ \\
\\
C: ${\rm CNOT}(1 \rightarrow 2)$ & $\overline{X}_1$ & $I \otimes X$ \\
& $\overline{X}_2$ & $X \otimes I$ \\
& $\overline{Z}_1$ & $I \otimes Z$ \\
& $\overline{Z}_2$ & $Z \otimes I$
\end{tabular}
\end{tabular}
\caption{Alice's quantum computer: a) network, b) analysis.}
\label{fig:swap}
\end{figure}

To figure this out, we will follow the evolution of four operators:
$\overline{X}_1$, $\overline{X}_2$, $\overline{Z}_1$, and
$\overline{Z}_2$.  The bars above the operators indicate that these
operators represent the {\em logical} operators $X_1$, $X_2$, $Z_1$,
and $Z_2$ from the beginning of the computation.  This terminology
will make it easier to follow their evolution through the complete
circuit.

Consider first $\overline{X}_1$.  It begins the computation as $X_1 =
X \otimes I$.  After the first CNOT (A), it becomes $X \otimes X$, as
per table~\ref{table:Clifford}.  We can rewrite this as
\begin{equation}
\left( X \otimes I \right) \left( I \otimes X \right),
\end{equation}
so step B (${\rm CNOT}(2 \rightarrow 1)$) maps $\overline{X}_1$ to
\begin{equation}
\left( X \otimes I \right) \left( X \otimes X \right) = I \otimes X.
\end{equation}
Then after step C, we still have $I \otimes X$, so the full circuit
maps
\begin{equation}
\overline{X}_1 = X \otimes I \rightarrow I \otimes X.
\end{equation}

We can perform a similar computation for $\overline{X}_2$,
$\overline{Z}_1$, and $\overline{Z}_2$.  The complete calculation is
summarized in figure~\ref{fig:swap}b.  The network of figure~\ref{fig:swap}a 
exchanges $X_1 \leftrightarrow X_2$ and $Z_1 \leftrightarrow Z_2$.  Therefore, 
it swaps the first and second qubits.

\begin{example}
The lever on Bob's quantum computer is stuck in the ``forward''
position, so it can only perform controlled-NOTs from qubit 1 to qubit
2.  He can still perform single-qubit operations normally.  How can he
perform a CNOT from qubit 2 to qubit 1?
\end{example}

\begin{figure}
\centering
\begin{tabular}{cc}
\begin{picture}(120,10)(0,60)

\put(0,60){a)}
\put(0,34){\makebox(40,12){$|\alpha\rangle$}}
\put(0,14){\makebox(40,12){$|\beta\rangle$}}

\put(40,40){\line(1,0){14}}
\put(66,40){\line(1,0){28}}
\put(106,40){\line(1,0){14}}
\put(40,20){\line(1,0){14}}
\put(66,20){\line(1,0){28}}
\put(106,20){\line(1,0){14}}

\put(54,34){\framebox(12,12){R}}
\put(54,14){\framebox(12,12){R}}

\put(80,40){\circle*{4}}
\put(80,40){\line(0,-1){24}}
\put(80,20){\circle{8}}

\put(94,34){\framebox(12,12){R}}
\put(94,14){\framebox(12,12){R}}

\put(54,48){\makebox(12,12){A}}
\put(74,48){\makebox(12,12){B}}
\put(94,48){\makebox(12,12){C}}

\end{picture}
& b)
\begin{tabular}[t]{lcc}
A: $R(1) R(2)$ & $\overline{X}_1$ & $Z \otimes I$ \\
& $\overline{X}_2$ & $I \otimes Z$ \\
& $\overline{Z}_1$ & $X \otimes I$ \\
& $\overline{Z}_2$ & $I \otimes X$ \\
\\
B: ${\rm CNOT}(1 \rightarrow 2)$ & $\overline{X}_1$ & $Z \otimes I$ \\
& $\overline{X}_2$ & $Z \otimes Z$ \\
& $\overline{Z}_1$ & $X \otimes X$ \\
& $\overline{Z}_2$ & $I \otimes X$ \\
\\
C: $R(1) R(2)$ & $\overline{X}_1$ & $X \otimes I$ \\
& $\overline{X}_2$ & $X \otimes X$ \\
& $\overline{Z}_1$ & $Z \otimes Z$ \\
& $\overline{Z}_2$ & $I \otimes Z$
\end{tabular}
\end{tabular}
\caption{Bob's ${\rm CNOT}(2 \rightarrow 1)$: a) network, b) analysis.}
\label{fig:CNOT}
\end{figure}
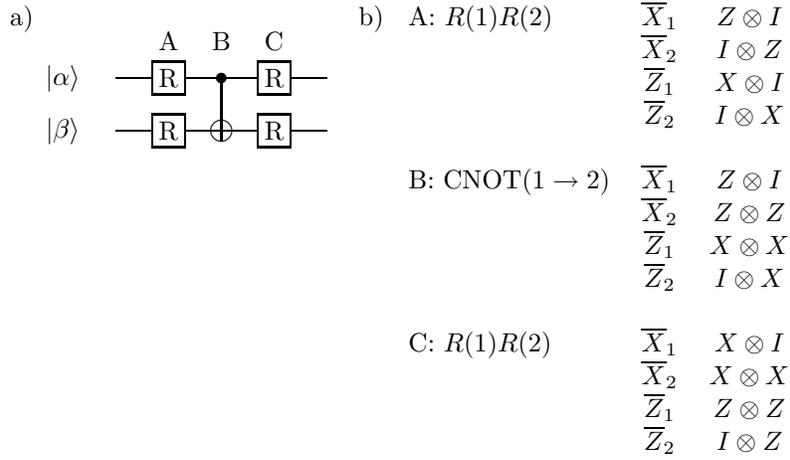

To solve this problem, the key is to notice the symmetric behavior of
the CNOT gate.  If we switch $X$ and $Z$ and qubits 1 and 2, we get
back the original transformation.  Therefore, the circuit of
figure~\ref{fig:CNOT}a acts as a ${\rm CNOT} (2 \rightarrow 1)$.
An analysis of the circuit is given in figure~\ref{fig:CNOT}b.
We recognize the final result as ${\rm CNOT}(2 \rightarrow 1)$, and so
conclude that the given circuit does produce the desired gate.

\begin{figure}
\centering
\begin{picture}(100,60)

\put(-20,60){a)}
\put(0,40){\line(1,0){14}}
\put(26,40){\line(1,0){74}}
\put(0,20){\line(1,0){14}}
\put(26,20){\line(1,0){28}}
\put(66,20){\line(1,0){34}}

\put(14,34){\framebox(12,12){R}}
\put(14,14){\framebox(12,12){P}}

\put(40,40){\circle*{4}}
\put(40,40){\line(0,-1){24}}
\put(40,20){\circle{8}}

\put(54,14){\framebox(12,12){R}}

\put(80,40){\circle*{4}}
\put(80,40){\line(0,-1){24}}
\put(80,20){\circle{8}}

\put(14,48){\makebox(12,12){A}}
\put(34,48){\makebox(12,12){B}}
\put(54,48){\makebox(12,12){C}}
\put(74,48){\makebox(12,12){D}}

\end{picture}

b) \begin{tabular}[t]{lcc|lcc}
A: $R(1) P(2)$ & $\overline{X}_1$ & $Z \otimes I$ &
C: $R(2)$ & $\overline{X}_1$ & $Z \otimes I$ \\
& $\overline{X}_2$ & $I \otimes Y$ &
& $\overline{X}_2$ & $- Z \otimes Y$ \\
& $\overline{Z}_1$ & $X \otimes I$ &
& $\overline{Z}_1$ & $X \otimes Z$ \\
& $\overline{Z}_2$ & $I \otimes Z$ &
& $\overline{Z}_2$ & $Z \otimes X$ \\
&&&&\\
B: ${\rm CNOT}(1 \rightarrow 2)$ & $\overline{X}_1$ & $Z \otimes I$ &
D: ${\rm CNOT}(1 \rightarrow 2)$ & $\overline{X}_1$ & $Z \otimes I$ \\
& $\overline{X}_2$ & $Z \otimes Y$ &
& $\overline{X}_2$ & $- I \otimes Y$ \\
& $\overline{Z}_1$ & $X \otimes X$ &
& $\overline{Z}_1$ & $- Y \otimes Y$ \\
& $\overline{Z}_2$ & $Z \otimes Z$ &
& $\overline{Z}_2$ & $Z \otimes X$
\end{tabular}
\caption{Ancient Hittite gate network: a) network, b) analysis}
\label{fig:Hittite}
\end{figure}
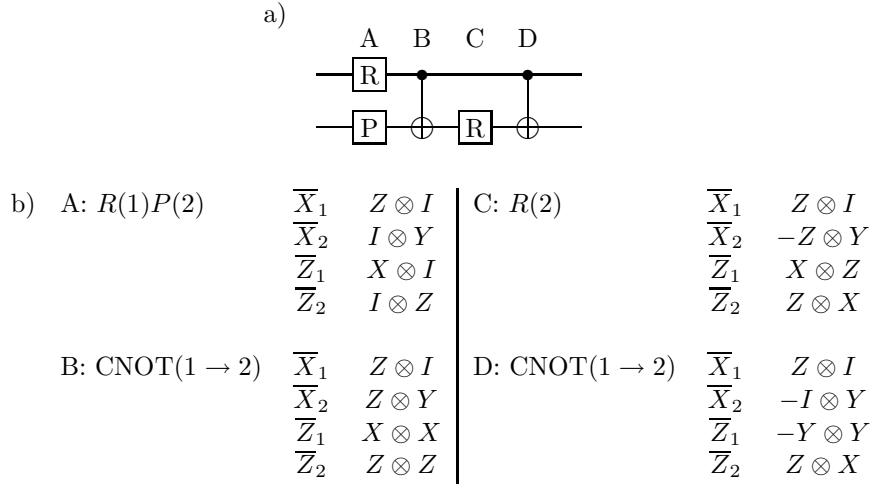

\begin{example}
Bob is attempting to read a paper written in ancient Hittite.  All he
can make sense of is a gate network, given in
figure~\ref{fig:Hittite}a.  What does this network do?
\end{example}

An analysis is given in figure~\ref{fig:Hittite}b.  Notice the two minus 
signs that appear.  For instance, in step C, for $\overline{X}_2$, we 
must apply $R$ to $Y$.  But
\begin{equation}
R (Y) = R (i X Z) = i R(X) R(Z) = i Z X = - Y.
\end{equation}
All in all, after the complete network, we get the transformation given 
for step D.

To convert back to the ket formalism, we can follow basis states.  For
instance, the initial state $|00\rangle$ starts as an eigenvector of
$Z \otimes I$ and $I \otimes Z$, with both eigenvalues $+1$.
Therefore, after the network, it will still be the +1 eigenvector of
both $\overline{Z}_1 = - Y \otimes Y$ and $\overline{Z}_2 = Z \otimes
X$.  Thus, we deduce that this network maps
\begin{equation}
|00\rangle \rightarrow \frac{1}{2} \left(|00\rangle + |01\rangle -
|10\rangle + |11\rangle \right).
\end{equation}

In addition, $|01\rangle = (I \otimes X)\,|00\rangle$, so $|01\rangle$
will map to $\overline{X}_2$ applied to the image of $|00\rangle$.
Since $\overline{X}_2 \rightarrow - I \otimes Y$,
\begin{equation}
|01\rangle \rightarrow \frac{i}{2} \left(-|01\rangle + |00\rangle +
|11\rangle + |10\rangle \right).
\end{equation}
We can similarly find the images of $|10\rangle$ and $|11\rangle$.  We
can put it all together into the following unitary matrix:
\begin{equation}
U = \frac{1}{2} \pmatrix{
\ 1 & \ i & \ 1 & \ i \cr
\ 1 &  -i & \ 1 &  -i \cr
 -1 & \ i & \ 1 &  -i \cr
\ 1 & \ i &  -1 &  -i}.
\end{equation}
This does not correspond to any well-known operation, so just what the
Hittites used it for must remain a mystery.

\section{Stabilizers}

\begin{example}
By dint of no little hard work, Alice has partially fixed her quantum
computer.  Now it only does 2 CNOTs at a time.  Unfortunately, she
can only get this improvement if she puts in a $|0\rangle$ as the
second input qubit (see figure~\ref{fig:XORswap}a).  What does it do now?
\end{example}
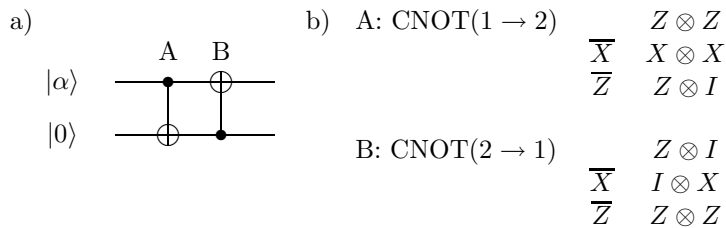
\begin{figure}
\centering
\begin{tabular}{cc}
\begin{picture}(100,10)(0,60)

\put(0,60){a)}
\put(0,34){\makebox(40,12){$|\alpha\rangle$}}
\put(0,14){\makebox(40,12){$|0\rangle$}}

\put(40,40){\line(1,0){60}}
\put(40,20){\line(1,0){60}}

\put(60,40){\circle*{4}}
\put(60,40){\line(0,-1){24}}
\put(60,20){\circle{8}}

\put(80,20){\circle*{4}}
\put(80,20){\line(0,1){24}}
\put(80,40){\circle{8}}

\put(54,46){\makebox(12,12){A}}
\put(74,46){\makebox(12,12){B}}

\end{picture}
& b)
\begin{tabular}[t]{lcc}
A: ${\rm CNOT}(1 \rightarrow 2)$ & & $Z \otimes Z$ \\
& $\overline{X}$ & $X \otimes X$ \\
& $\overline{Z}$ & $Z \otimes I$ \\
\\
B: ${\rm CNOT}(2 \rightarrow 1)$ & & $Z \otimes I$ \\
& $\overline{X}$ & $I \otimes X$ \\
& $\overline{Z}$ & $Z \otimes Z$
\end{tabular}
\end{tabular}
\caption{Alice's improved quantum computer: a) network, b) analysis.}
\label{fig:XORswap}
\end{figure}

The input state for this network will always be a $+1$ eigenvector of
$I \otimes Z$.  Therefore, the state of the system will always be a
$+1$ eigenvector of $\overline{Z}_2$.  Since the eigenvalue of
$\overline{Z}_2$ is fixed, there is no point in following the
evolution of $\overline{X}_2$ as well, since it would act to move the
state to the irrelevant $-1$ eigenvector of $\overline{Z}_2$.  The
full analysis of this circuit is given in figure~\ref{fig:XORswap}b.

The format of this analysis is slightly different.  The unlabelled
operators are those for which the state of the computer is a $+1$
eigenvector --- in this case, $\overline{Z}_2$.  In addition, since
among the labelled operators, we are only following $\overline{X}_1$
and $\overline{Z}_1$, they are instead labelled simply $\overline{X}$
and $\overline{Z}$.

The final result may not be instantly recognizable.  We can see that
the state is in a $+1$ eigenvector of $Z \otimes I$, so the first
qubit is $|0\rangle$.  Since $Z \otimes I$ acts as the identity on all
possible final states of the computer, it also follows that
$\overline{Z} = Z \otimes Z$ is equivalent to $\left(Z \otimes I
\right) \left(Z \otimes Z \right) = I \otimes Z$.  Therefore,
\begin{eqnarray}
\overline{X} & \rightarrow & I \otimes X, \\
\overline{Z} & \rightarrow & I \otimes Z.
\end{eqnarray}
In other words, the original first qubit has migrated to the second.
Alice's computer still performs a swap of the first and second qubits.

In the previous example, only a single qubit had a fixed input value,
but more generally, multiple qubits may be fixed or constrained.  As
above, it will be helpful to consider the set of operators in ${\cal
P}$ for which the input states are $+1$ eigenvectors.  The set of such
operators is closed under multiplication, and therefore forms a group,
known as the {\em stabilizer} $S$~\cite{gottesman:stab, CRSS:stab}.  
\begin{equation}
S = \left\{ M \in {\cal P} \mbox{ such that } M |\psi\rangle =
|\psi\rangle \mbox{ for all allowed inputs } |\psi\rangle \right\}
\end{equation}
In the previous example, the stabilizer only contained two elements:
$I \otimes Z$ and the identity $I \otimes I$.

The stabilizer is always an Abelian group: If $M, N \in S$, then
\begin{eqnarray}
M N |\psi\rangle & = & |\psi\rangle \\
N M |\psi\rangle & = & |\psi\rangle \\
{}[M, N] |\psi\rangle & = & 0.
\end{eqnarray}
Any two elements of ${\cal P}$ either commute or anticommute, so it
follows that $M$ and $N$ commute.  In addition, all elements of ${\cal
P}$ square to $\pm 1$ (operators with an overall phase of $\pm 1$
square to $+1$, operators with an overall phase of $\pm i$ square to
$-1$).  The operators that square to $-1$ have imaginary eigenvalues,
while the operators that square to $+1$ have eigenvalues $\pm 1$.
Therefore, any stabilizer is composed of operators in ${\cal P}$ which
have an overall phase of $\pm 1$, and which square to $+1$.  It
follows that $S$ is isomorphic to $({\bf Z}_2)^k$ for some $k$.

In fact, $k$ is exactly the number of input qubits that are fixed.
The requirement that a single qubit is fixed provides a single
operator for the stabilizer.  The $k$ operators together then generate
$S$ --- the $2^k$ elements of $S$ are products of the generators.

The Pauli group operators that remain interesting are the operators
that act on the other $n-k$ qubits (if there are $n$ total qubits).
There are a total of $2(n-k)$ such independent operators.  When the
stabilizer puts more complicated constraints on the state, the
$\overline{X}$ and $\overline{Z}$ operators are any $2(n-k)$ operators
that commute with $S$.

\begin{figure}
\centering
\begin{tabular}{cc}
\begin{picture}(100,10)(0,60)

\put(0,60){a)}
\put(0,34){\makebox(40,12){$|0\rangle$}}
\put(0,14){\makebox(40,12){$|0\rangle$}}

\put(40,40){\line(1,0){14}}
\put(66,40){\line(1,0){34}}
\put(40,20){\line(1,0){60}}

\put(54,34){\framebox(12,12){R}}
\put(80,40){\circle*{4}}
\put(80,40){\line(0,-1){24}}
\put(80,20){\circle{8}}

\put(54,48){\makebox(12,12){A}}
\put(74,48){\makebox(12,12){B}}

\end{picture}
& b)
\begin{tabular}[t]{lc}
Start & $Z \otimes I$ \\
& $I \otimes Z$ \\
\\
A: $R(1)$ & $X \otimes I$ \\
& $I \otimes Z$ \\
\\
B: ${\rm CNOT}(1 \rightarrow 2)$ & $X \otimes X$ \\
& $Z \otimes Z$
\end{tabular}
\end{tabular}
\caption{Making a Bell state: a) network, b) analysis.}
\label{fig:EPR}
\end{figure}
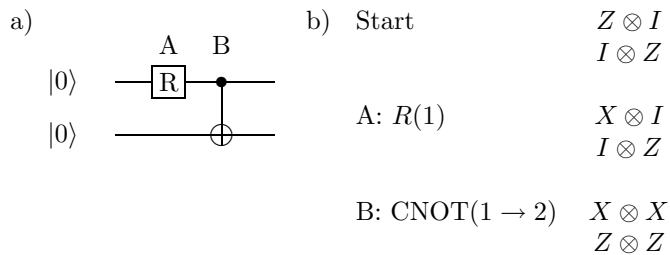

\begin{example}
Alice's quantum computer is finally fixed.  She wishes to demonstrate
her devotion to Bob by sending him half of a Bell state, along with a
card reading ``forever entagled.''  She performs the network of
figure~\ref{fig:EPR}a and sends the card and second qubit to Bob.
\end{example}

The analysis of the network appears in figure~\ref{fig:EPR}b.
In this case, we only need to consider elements of the stabilizer.
Again, the generators of the stabilizer are unlabelled.  Their order
does not matter --- we could switch the generators, and they still
generate the same group.  In fact, we could exchange one or more of
the generators with other elements of $S$ without affecting anything,
as long as we still have a set of $n$ independent operators to act as
generators.  The transformation of the rest of the stabilizer is
completely determined by the transformation of the generators.

The final state in this case has stabilizer $\{I \otimes I,\ X \otimes
X,\ - Y \otimes Y,\ Z \otimes Z \}$.  There is a single state which
has eigenvalue $+1$ for all four of these operators.  It is given (up
to normalization) by
\begin{equation}
|\psi\rangle = \left( \sum_{M \in S} M \right) |00\rangle.
\end{equation}
This is the correct state because acting on it with $N \in S$ just
permutes the terms in the sum, giving the same state.\footnote{If $- Z
\otimes Z$ had been in the stabilizer instead of $Z \otimes Z$,
$|\psi\rangle$ would have been null.  To produce the correct state, we
would have had to start with some state other than $|00\rangle$ on the
right.}  In this case,
the state is
\begin{equation}
\frac{1}{\sqrt{2}} \left( |00\rangle + |11\rangle \right),
\end{equation}
which is indeed a Bell state.

A state which can be completely described by specifying the stabilizer
is called a {\em stabilizer state}.  There are many states in the
Hilbert space which are not stabilizer states (in fact, there are only
a finite number of stabilizer states of a given size), but many of the
most interesting states are stabilizer states.

\begin{example}
Alice decides she would rather send the singlet state, but doesn't
want to go through the bother of another CNOT.  How can she produce
the singlet from what she currently has?
\end{example}

The singlet state is
\begin{equation}
\frac{1}{\sqrt{2}} \left( |01\rangle - |10\rangle \right).
\end{equation}
It has the stabilizer $\{I \otimes I,\ - X \otimes X,\ - Y \otimes Y,\
- Z \otimes Z\}$ (generated by $- X \otimes X$ and $- Z \otimes Z$).
This is clearly very similar to her current stabilizer (generated by
$X \otimes X$ and $Z \otimes Z$).  All she needs to do is apply an
operator that maps $X \otimes X \rightarrow -X \otimes X$ and $Z
\otimes Z\rightarrow -Z \otimes Z$.

One possible operator is $Y \otimes I$.
\begin{eqnarray}
Y (X) = Y X Y^\dagger & = & - Y Y^\dagger X = - X, \\
Y (Z) = Y Z Y^\dagger & = & - Y Y^\dagger Z = - Z.
\end{eqnarray}
Another operator could be $I \otimes Y$.  $X \otimes Z$ would also
work:
\begin{eqnarray}
\left(X \otimes Z\right) \left(X \otimes X\right) \left(X \otimes
Z\right) & = & - X \otimes X, \\
\left(X \otimes Z\right) \left(Z \otimes Z\right) \left(X \otimes
Z\right) & = & - Z \otimes Z.
\end{eqnarray}

In fact, any operator $E \in {\cal P}$ that anticommutes with both $X
\otimes X$ and $Z \otimes Z$ would work: if $\{E, M\} = 0$, then
\begin{equation}
E (M) = E M E^\dagger = - E E^\dagger M = - M. \label{anticommute}
\end{equation}

\section{Quantum Error-Correcting Codes}

\begin{example}
Alice and Bob sometimes bring quantum data home with them.  Their
two-year-old child Alice, Jr.\ is very curious.  Even if the data is
stored on a high shelf, Alice, Jr.\ sometimes gets to it.  When she
does, she puts a single qubit in her mouth, randomizing it, then puts
it back.  After that, she loses interest and goes off to create
trouble elsewhere.  How can Alice and Bob tell if Alice, Jr.\ has
disturbed their state?  They do not want to destroy it if the state is
OK, but they are willing to discard it if it has been changed.
\end{example}

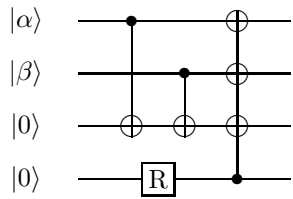
\begin{figure}
\centering
\begin{picture}(120,100)

\put(0,74){\makebox(40,12){$|\alpha\rangle$}}
\put(0,54){\makebox(40,12){$|\beta\rangle$}}
\put(0,34){\makebox(40,12){$|0\rangle$}}
\put(0,14){\makebox(40,12){$|0\rangle$}}

\put(40,80){\line(1,0){80}}
\put(40,60){\line(1,0){80}}
\put(40,40){\line(1,0){80}}
\put(40,20){\line(1,0){24}}
\put(76,20){\line(1,0){44}}

\put(60,80){\circle*{4}}
\put(60,80){\line(0,-1){44}}
\put(60,40){\circle{8}}

\put(64,14){\framebox(12,12){R}}

\put(80,60){\circle*{4}}
\put(80,60){\line(0,-1){24}}
\put(80,40){\circle{8}}

\put(100,20){\circle*{4}}
\put(100,20){\line(0,1){64}}
\put(100,40){\circle{8}}
\put(100,60){\circle{8}}
\put(100,80){\circle{8}}

\end{picture}
\caption{A single qubit error-detecting code}
\label{fig:fourqubit}
\end{figure}

Alice and Bob cannot simply measure their quantum state, since that
would collapse any superposition it might be in.  Instead, the
solution is to group the qubits in sets of two.  To each pair, they add
two qubits in a known state, and apply some operations resulting in
the stabilizer $S$ generated by 
\begin{eqnarray}
X \otimes X \otimes X \otimes X \nonumber \\
Z \otimes Z \otimes Z \otimes Z. \label{fourqubit}
\end{eqnarray}
For instance, they could use the
circuit in figure~\ref{fig:fourqubit}.

The stabilizer~(\ref{fourqubit}) has the advantage that any
single-qubit operator anticommutes with at least one of the
generators.  Therefore, if we apply any error consisting of a
single-qubit operator $E \in {\cal P}$, at least one of the generators
will change sign, as per~(\ref{anticommute}).  We can therefore detect
that $E$ has occurred by measuring the generators of $S$.  If the
eigenvalues are $+1$, no error has occurred (or two or more errors
have occurred).

Of course, randomizing a qubit would generally not correspond to
performing an operator in ${\cal P}$.  However, ${\cal P}$ does span
the set of matrices, so the most general error will be the sum of
operators from ${\cal P}$.  Such an error would put the state in a
superposition of eigenstates of $S$, but measuring the generators of
$S$ collapses the state into just one eigenstate.  This has the effect
of collapsing the error into some operator from ${\cal P}$.

The states described by~(\ref{fourqubit}) form a {\em quantum
error-detecting code} with the ability to detect one
error~\cite{detect,erasure}.  A code that detects any error affecting
less than $d$ qubits is said to have distance $d$.  The four-qubit
code thus has distance 2.  Because the code is completely determined
by its stabilizer, it is said to be a {\em stabilizer code}.

\begin{example}
Alice and Bob wish to bring some particularly important quantum data
home.  They worked very hard to produce it, and do not want to have to
discard it if Alice, Jr.\ ruins a single qubit.  What can they do?
\end{example}

Now, Alice and Bob have to go beyond an error-detecting code to a {\em
quantum error-correcting code}.  However, the principles are the
same.  In order to correct a single error, it is sufficient for Alice
and Bob to be able to figure out exactly what the error was.  Since it
is from ${\cal P}$ (or has been collapsed to an error from ${\cal
P}$), they can then just apply the inverse to restore the original
state.

Distinguishing the state $E |\psi\rangle$ from $F |\psi\rangle$ is the
same problem as distinguishing the state $E F
|\psi\rangle$\footnote{Or more precisely, $E^\dagger F |\psi\rangle$,
which is the same for Hermitian $E$.} from the state $|\psi\rangle$.
Therefore, a code that detects the error $E F$ will distinguish the
errors $E$ and $F$.  In other words, to correct any single-qubit
error, the code must detect all two-qubit errors.  More generally, to
correct $t$ arbitrary errors, the code should have distance $2t + 1$.

The stabilizer of the smallest error-correcting code to fix one error
is given in table~\ref{table:fivequbit}~\cite{LANL:5qubit,IBM:5qubit}.
\begin{table}
\centering
\begin{tabular}{lccccc}
& $X$ & $Z$ & $Z$ & $X$ & $I$ \\
& $I$ & $X$ & $Z$ & $Z$ & $X$ \\
& $X$ & $I$ & $X$ & $Z$ & $Z$ \\
& $Z$ & $X$ & $I$ & $X$ & $Z$ \\
\\
$\overline{X}$ & $X$ & $X$ & $X$ & $X$ & $X$ \\
$\overline{Z}$ & $Z$ & $Z$ & $Z$ & $Z$ & $Z$
\end{tabular}
\caption{The five-qubit code}
\label{table:fivequbit}
\end{table}
It encodes a single qubit in five qubits, and has distance three.  We
say it is a $[5, 1, 3]$ quantum code (or a $[[5, 1, 3]]$ code).

The $\overline{X}$ and $\overline{Z}$ operators are, as before, the
logical operators corresponding to operations on the original data
qubits.  By analyzing their behavior, we could, for instance,
understand how to perform operations on the data without first
decoding the state ({\em fault-tolerant} operations
\cite{gottesman:FT,shor:FT}).

I said above that a stabilizer code has distance $d$ when for every
Pauli group operator $E$ with weight less than $d$, there is an
element $M$ of the stabilizer that anticommutes with $E$.  In fact,
the requirement is slightly weaker.  It is possible for $E
|\psi\rangle = |\psi\rangle$ for all codewords $|\psi\rangle$ for some
particular $E$.  In this case, there is no way the code will be able
to detect that $E$ has occurred --- but there is no need to do so,
since the state produced is exactly the correct state.  If $E
|\psi\rangle = |\psi\rangle\ \forall |\psi\rangle$, then $E \in S$.
Therefore, the complete condition for a stabilizer code to have
distance $d$ is that for all $E \in {\cal P}$ of weight less than $d$,
either $\exists M \in S$ with $\{M, E\} = 0$, or $E \in S$.  If the
second condition is ever needed, the code is said to be {\em degenerate}.
If the second condition is not used, as for the four- and five-qubit
codes above, the code is {\em nondegenerate}.

Quantum error-correcting codes solve a serious difficulty in the
design of quantum computers.  Simply having objects that behave in a
more-or-less quantum fashion is insufficient to produce a quantum
computer.  If a qubit interacts with its environment, it will have a
tendency to act as if it is in one or the other basis state (for some
particular basis, which depends on the interaction), just as if it had
been measured.  A quantum computer with strong interactions with the
environment will thus tend to act as if it is in one of the $2^N$
basis states --- in other words, it will act just like a classical
computer.  Using quantum error correction and fault-tolerant
operations, we can greatly simplify our task: instead of having to
build a quantum computer with essentially no uncontrolled interaction
with the environment, we need only build one where the interactions
are small enough to give us time to do
error-correction.\footnote{Only.  Can you tell I'm a theorist?}

\section{Measurements}

\begin{example}
Bob has bumped his head and cannot remember how to perform the P
gate.  Luckily, he does remember how to perform the CNOT, the Pauli
group, and how to measure operators in the Pauli group.  How can he
make a P gate?
\end{example}

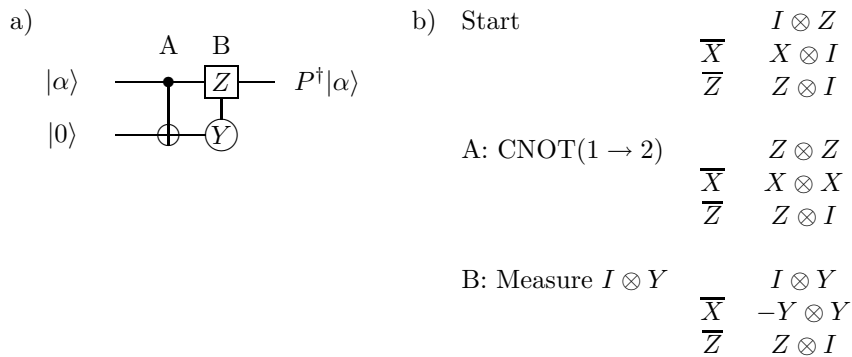
\begin{figure}
\centering
\begin{tabular}{cc}
\begin{picture}(140,10)(0,60)

\put(0,60){a)}
\put(0,34){\makebox(40,12){$|\alpha\rangle$}}
\put(0,14){\makebox(40,12){$|0\rangle$}}
\put(100,34){\makebox(40,12){$P^\dagger |\alpha\rangle$}}

\put(40,40){\line(1,0){34}}
\put(86,40){\line(1,0){14}}
\put(40,20){\line(1,0){34}}

\put(60,40){\circle*{4}}
\put(60,40){\line(0,-1){24}}
\put(60,20){\circle{8}}

\put(80,20){\circle{12}}
\put(74,14){\makebox(12,12){$Y$}}
\put(80,26){\line(0,1){8}}
\put(74,34){\framebox(12,12){$Z$}}

\put(54,48){\makebox(12,12){A}}
\put(74,48){\makebox(12,12){B}}

\end{picture}
& b)
\begin{tabular}[t]{lcc}
Start& & $I \otimes Z$ \\
& $\overline{X}$ & $X \otimes I$ \\
& $\overline{Z}$ & $Z \otimes I$ \\
\\
A: ${\rm CNOT(1 \rightarrow 2)}$ & & $Z \otimes Z$ \\
& $\overline{X}$ & $X \otimes X$ \\
& $\overline{Z}$ & $Z \otimes I$ \\
\\
B: Measure $I \otimes Y$ & & $I \otimes Y$ \\
& $\overline{X}$ & $- Y \otimes Y$ \\
& $\overline{Z}$ & $Z \otimes I$ 
\end{tabular}
\end{tabular}
\caption{Creating the P gate: a) network, b) analysis.}
\label{fig:Pgate}
\end{figure}

The solution is given in figure~\ref{fig:Pgate}a.
A full analysis appears in figure~\ref{fig:Pgate}b.
If the initial state $|\alpha\rangle = a|0\rangle + b|1\rangle$, then
the state after completing step A is
\begin{equation}
a|00\rangle + b|11\rangle = \frac{1}{2} \left( a|0\rangle -
ib|1\rangle \right) \left( |0\rangle + i |1\rangle \right) +
\frac{1}{2} \left( a|0\rangle + ib|1\rangle \right) \left( |0\rangle -
i |1\rangle \right).
\end{equation}
Therefore, measuring $I \otimes Y$ will yield either the state
\begin{equation}
\left( a|0\rangle - ib|1\rangle \right) \left( |0\rangle + i |1\rangle
\right),
\end{equation}
or the state
\begin{equation}
\left( a|0\rangle + ib|1\rangle \right) \left( |0\rangle - i |1\rangle
\right).
\end{equation}
The first has stabilizer $I \otimes Y$ and $\overline{X} = - Y \otimes
I$, while the second has stabilizer $- I \otimes Y$ and $\overline{X} =
+ Y \otimes I$.  In both cases, $\overline{Z} = Z \otimes I$.  The two
possible states after the measurement are related by the action of the
operator $Z \otimes Z$.  Note that $Z \otimes Z$ is the stabilizer
after step A.

Therefore, step B consists of measuring $I \otimes Y$ and performing
$Z \otimes Z$ if the result is $-1$.  That produces the final state
given in the table.  Since $I \otimes Y$ is in the stabilizer, $Y
\otimes Y$ is equivalent to $Y \otimes I$ on the states of interest.
We can recognize the final operation as a $P^\dagger$ gate.  We can get 
a regular P gate either by performing this 3 times, or more easily by
applying $Z$ once to the output qubit.

The fact that we were able to convert the $+1$ and $-1$ measurement
results into the same state was no fluke.  Generally, suppose we have
a state which is partially (or completely) described by a stabilizer $S$.
Suppose we wish to measure an operator $A \in {\cal P}$ which
anticommutes with some element $M \in S$.  The measurement performs
one of the operators
\begin{equation}
P_{\pm} = \frac{1}{2} (I \pm A),
\end{equation}
depending on whether the measurement result is $\pm 1$.  But
\begin{equation}
M P_- M^\dagger = \frac{1}{2} M (I - A) M^\dagger = \frac{1}{2} (I +
A) M M^\dagger = P_+.
\end{equation}
Therefore, by applying $M$ whenever the measurement result is $-1$, we
can ensure that we always get the state $P_+ |\psi\rangle$.

We can go further and describe the evolution of the state using the
Heisenberg formalism.  The state $P_+ |\psi\rangle$ will always be in
a $+1$ eigenstate of $A$, so $A$ is a member of the new stabilizer
$S'$.  $M$, on the other hand, will not be in $S'$.  The other
generators will be in $S'$ if they commute with $A$, since commuting
observables can have simultaneous eigenvectors.  On the other hand, if
$N \in S$ does not commute with $A$, $N \notin S'$; but $M N$ does
commute with $A$, so $M N \in S'$.

We can also follow the evolution of the $\overline{X}$ and
$\overline{Z}$ operators.  If they commute with $A$, they are
unaffected by the measurement of $A$.  Our original presentation of
$\overline{X}_j$ or $\overline{Z}_j$ might not commute with $A$, but
recall that operators acting on the state are equivalent up to
multiplication by elements of the stabilizer.  Therefore, even if
$\overline{X}_j$ does not commute with $A$, $M \overline{X}_j$ does,
and acts the same way.  Therefore, under the measurement,
$\overline{X}_j \rightarrow M \overline{X}_j$.

We can sum up the rules for evolving the operators after a measurement
of $A$ (and correction if the result is $-1$) as follows:
\begin{enumerate}
\item Identify $M \in S$ satisfying $\{M, A\} = 0$.
\item Remove $M$ from the stabilizer
\item Add $A$ to the stabilizer
\item For each $N$, where $N$ runs over the other generators of $S$
and the $\overline{X}$ and $\overline{Z}$ operators, leave $N$ alone
if $[N, A] = 0$, and replace $N$ with $MN$ if $\{N, A\} = 0$.
\end{enumerate}

\begin{example}[Quantum Teleportation]
Alice needs to quickly send a qubit to Bob, but Quantum Parcel
Services (QPS) is on strike.  Luckily, Alice shares an EPR pair with
Bob, and the regular classical phone lines are still open.  How can
she get him the qubit?
\end{example}

The solution to this problem is quantum
teleportation~\cite{teleport}.  It is straightforward to analyze in
the Heisenberg representation.  I will assume Alice and Bob start with
the Bell state $|00\rangle + |11\rangle$.  The network is given in
figure~\ref{fig:teleport}a.
\begin{figure}
\centering
\begin{picture}(160, 100)

\put(0,100){a)}
\put(0,74){\makebox(40,12){$|\psi\rangle$}}
\put(120,14){\makebox(40,12){$|\psi\rangle$}}

\put(40,80){\line(1,0){34}}
\put(40,60){\line(1,0){34}}
\put(86,60){\line(1,0){8}}
\put(40,20){\line(1,0){34}}
\put(86,20){\line(1,0){8}}
\put(106,20){\line(1,0){14}}

\put(20,40){\line(1,1){20}}
\put(20,40){\line(1,-1){20}}

\put(60,80){\circle*{4}}
\put(60,80){\line(0,-1){24}}
\put(60,60){\circle{8}}

\put(80,80){\circle{12}}
\put(74,74){\makebox(12,12){$X$}}
\put(80,74){\line(0,-1){8}}
\put(74,54){\framebox(12,12){$Z$}}
\put(80,54){\line(0,-1){28}}
\put(74,14){\framebox(12,12){$Z$}}

\put(100,60){\circle{12}}
\put(94,54){\makebox(12,12){$Z$}}
\put(100,54){\line(0,-1){28}}
\put(94,14){\framebox(12,12){$X$}}

\put(54,88){\makebox(12,12){A}}
\put(74,88){\makebox(12,12){B}}
\put(94,88){\makebox(12,12){C}}

\end{picture}

b) \begin{tabular}[t]{lcccc}
Start & & $I$ & $X$ & $X$ \\
& & $I$ & $Z$ & $Z$ \\
& $\overline{X}$ & $X$ & $I$ & $I$ \\
& $\overline{Z}$ & $Z$ & $I$ & $I$ \\
\\
A: ${\rm CNOT}(1 \rightarrow 2)$ & & $I$ & $X$ & $X$ \\
& & $Z$ & $Z$ & $Z$ \\
& $\overline{X}$ & $X$ & $X$ & $I$ \\
& $\overline{Z}$ & $Z$ & $I$ & $I$ \\
\\
B: Measure $X \otimes I \otimes I$ & & & $X$ & $X$ \\
& $\overline{X}$ & & $X$ & $I$ \\
& $\overline{Z}$ & & $Z$ & $Z$ \\
\\
C: Measure $Z \otimes I$ & $\overline{X}$ & & & $X$ \\
& $\overline{Z}$ & & & $Z$
\end{tabular}
\caption{Quantum teleportation: a) network, b) analysis.}
\label{fig:teleport}
\end{figure}
Alice and Bob begin with one EPR pair.  Alice and Bob perform some
local operations, and in the process Alice sends Bob two classical
bits.  The net result is that Alice's qubit is sent to Bob.  The
Heisenberg representation analysis is given in
figure~\ref{fig:teleport}b.  The two classical bits sent from Alice
to Bob are the two measurement results, which are required for Bob
to perform the correct operations to turn the state into the +1
eigenvector of the two measured operators.

Since in this network, once a qubit is measured, it is completely
determined and never reused, we drop measured qubits from the
analysis.  A more complicated network might include partial entangled
measurements, in which case the measured qubits would still have some
interesting remaining degrees of freedom and should be retained in the
description.  Note that when dropping qubits, the size of the
stabilizer shrinks, but we keep the same number of $\overline{X}$ and
$\overline{Z}$ operators, since we only drop qubits which are
completely constrained by the stabilizer.

\begin{example}[Remote XOR]
Alice forgot to perform a CNOT from another qubit of hers to the one
she sent to Bob.  Unfortunately, they only have one more shared EPR
pair (Bob has been gambling at the entanglement casino and losing
heavily).  Can she still perform the CNOT?
\end{example}

The network that allows Alice to perform a remote XOR to Bob's qubit
is given in figure~\ref{fig:RXOR}a.
\begin{figure}
\centering
\begin{picture}(120,120)

\put(0,120){a)}
\put(40,100){\line(1,0){54}}
\put(40,80){\line(1,0){34}}
\put(40,40){\line(1,0){34}}
\put(86,40){\line(1,0){8}}
\put(40,20){\line(1,0){34}}
\put(86,20){\line(1,0){34}}

\put(20,60){\line(1,1){20}}
\put(20,60){\line(1,-1){20}}

\put(60,100){\circle*{4}}
\put(60,100){\line(0,-1){24}}
\put(60,80){\circle{8}}

\put(60,40){\circle*{4}}
\put(60,40){\line(0,-1){24}}
\put(60,20){\circle{8}}

\put(80,80){\circle{12}}
\put(74,74){\makebox(12,12){$Z$}}
\put(80,74){\line(0,-1){28}}
\put(74,34){\framebox(12,12){$X$}}
\put(80,34){\line(0,-1){8}}
\put(74,14){\framebox(12,12){$X$}}

\put(100,40){\circle{12}}
\put(94,34){\makebox(12,12){$X$}}
\put(100,46){\line(0,1){48}}
\put(94,94){\framebox(12,12){$Z$}}

\put(54,108){\makebox(12,12){A}}
\put(74,108){\makebox(12,12){B}}
\put(94,108){\makebox(12,12){C}}

\end{picture}

b) \begin{tabular}[t]{lccccc}
Start & & $I$ & $X$ & $X$ & $I$ \\
& & $I$ & $Z$ & $Z$ & $I$ \\
& $\overline{X}_A$ & $X$ & $I$ & $I$ & $I$ \\
& $\overline{Z}_A$ & $Z$ & $I$ & $I$ & $I$ \\
& $\overline{X}_B$ & $I$ & $I$ & $I$ & $X$ \\
& $\overline{Z}_B$ & $I$ & $I$ & $I$ & $Z$ \\
\\
A: ${\rm CNOT} (1 \rightarrow 2) {\rm CNOT} (3 \rightarrow 4)$ & & $I$
& $X$ & $X$ & $X$ \\
& & $Z$ & $Z$ & $Z$ & $I$ \\
& $\overline{X}_A$ & $X$ & $X$ & $I$ & $I$ \\
& $\overline{Z}_A$ & $Z$ & $I$ & $I$ & $I$ \\
& $\overline{X}_B$ & $I$ & $I$ & $I$ & $X$ \\
& $\overline{Z}_B$ & $I$ & $I$ & $Z$ & $Z$ \\
\\
B: Measure $I \otimes Z \otimes I \otimes I$ & & $Z$ & & $Z$ & $I$ \\
& $\overline{X}_A$ & $X$ & & $X$ & $X$ \\
& $\overline{Z}_A$ & $Z$ & & $I$ & $I$ \\
& $\overline{X}_B$ & $I$ & & $I$ & $X$ \\
& $\overline{Z}_B$ & $I$ & & $Z$ & $Z$ \\
\\
C: Measure $I \otimes X \otimes I$
& $\overline{X}_A$ & $X$ & & & $X$ \\
& $\overline{Z}_A$ & $Z$ & & & $I$ \\
& $\overline{X}_B$ & $I$ & & & $X$ \\
& $\overline{Z}_B$ & $Z$ & & & $Z$
\end{tabular}
\caption{The remote XOR: a) network, b) analysis.}
\label{fig:RXOR}
\end{figure}
Alice and Bob start with one EPR pair.  They each perform a CNOT and a
measurement, then send each other their measurement results (for a
total of one classical bit each direction).  The result is a CNOT from
Alice's qubit to Bob's.  This operation has also been studied by
\cite{RXOR}.  The analysis appears in figure~\ref{fig:RXOR}b.

\section{Discussion and Summary}

The methods comprising the Heisenberg representation of quantum
computation allow a much more rapid analysis of many networks than can
be done using the usual methods of multiplying together the full $2^n
\times 2^n$ unitary matrices describing the evolution of an $n$ qubit
system, or by directly following the evolution of state vectors in
such a system.  For most networks, the usual methods require keeping
track of an exponential number of matrix elements or coefficients.

In contrast, for a network composed only of gates from the Clifford
group and measurements of Pauli group operators, the Heisenberg
representation provides an excellent method of describing the system.
In such networks, it is only necessary to follow the evolution of at
most $2n$ $\overline{X}$ and $\overline{Z}$ operators, and each one is
a member of the Pauli group, so can be described by $2n + 1$ bits.
Thus, the time and space required to analyze such a network on a
classical computer are polynomials in $n$, instead of exponentials.

In fact, we can go beyond just Clifford group elements and Pauli group
measurements, and also add Clifford group operations conditioned on
classical bits (which might, of course, be the results of measurements
performed earlier in the computation).  This collection of
observations forms the proof of the following
theorem~\cite{knillsthm}:

\begin{theorem}[Knill's theorem]
Any quantum computer performing only: a) Clifford group gates, b)
measurements of Pauli group operators, and c) Clifford group operations
conditioned on classical bits, which may be the results of earlier
measurements, can be perfectly simulated in polynomial time on a
probabilistic classical computer.
\end{theorem}

Of course, Clifford group operations and Pauli group measurements do
not provide a universal set of quantum gates.  Another gate is needed,
which could be a quantum version of the classical Toffoli gate
($|a\rangle|b\rangle|c\rangle \rightarrow |a\rangle|b\rangle|c +
ab\rangle$), a $\pi/8$ rotation of the Bloch sphere, or the square
root of the controlled-NOT gate, among other possibilities.  The
Clifford group plus an appropriate extra gate generate a set of
unitary operators dense in $U(2^n)$ and therefore form a universal set
of gates.

Knill's theorem implies that quantum computation is only more powerful
than classical computation when it uses gates outside the Clifford
group.  However, networks using only Clifford group gates also have a
number of important applications in the area of quantum
communications.  Quantum error-correcting codes are an important
example --- stabilizer codes use only Clifford group gates for
encoding and decoding, yet are extremely useful for overcoming the
effects of errors and decoherence.  Other important communication
problems, such as quantum teleportation, also use only Clifford group
gates and measurements.  Finally, networks consisting of Clifford
group gates and measurements may provide useful subroutines to more
complex quantum computations; an efficient method of analyzing and
searching for such subroutines could prove very useful.

This preprint is Los Alamos preprint number LAUR-98-2848.

\end{document}